\documentclass[aps,prd,eqsecnum,nofootinbib,showpacs,twocolumn]{revtex4-1}

\usepackage{amsmath,amssymb,bm}
\usepackage[dvips]{graphicx}
\usepackage{color}

\newcommand{\be}{\begin{equation}}
\newcommand{\ee}{\end{equation}}
\newcommand{\bea}{\begin{eqnarray}}
\newcommand{\eea}{\end{eqnarray}}

\newcommand{\eq}[1]{Eq.~(\ref{eq:#1})}
\newcommand{\sect}[1]{Sec.~\ref{sec:#1}}
\newcommand{\appen}[1]{Appendix~\ref{sec:#1}}

\newcommand{\del}{\partial}

\newcommand{\eg}{{\it e.g.}}
\newcommand{\ie}{{\it i.e.}}

\newcommand{\Nfour}{${\cal N}=4$}

\bmdefine{\bmq}{{\bm{q}}}
\bmdefine{\bmk}{{\bm{k}}}
\bmdefine{\bmx}{{\bm{x}}}
\bmdefine{\bmy}{{\bm{y}}}
\bmdefine{\bmr}{{\bm{r}}}
\bmdefine{\bmnabla}{{\bm{\nabla}}}
\bmdefine{\bmA}{ \bm{A} }
\bmdefine{\bmD}{ \bm{D} }

\bmdefine{\bmPhi}{ \bm{\Phi} }
\bmdefine{\bmPsi}{ \bm{\Psi} }
\bmdefine{\bmcalO}{ \bm{\mathcal{O}} }

\newcommand{\calL}{{\cal L}}
\newcommand{\calM}{{\cal M}}

\newcommand{\vecx}{\vec{x}}

\newcommand{\nq}{\mathfrak{q}}

\newcommand{\nw}{\mathfrak{w}}

\bmdefine{\bmg}{{\bm{g}}}
\bmdefine{\bmR}{{\bm{R}}}

\newcommand{\mfh}{\mathfrak{h}}

\newcommand{\mfA}{\mathfrak{A}}

\newcommand{\dw}{\delta\nw}
\newcommand{\dq}{\delta \nq}
\newcommand{\dqs}{\delta(\nq^2)}

\newcommand{\NGB}{N_\text{\tiny{GB}}}
\newcommand{\qN}{\nq_N}
\newcommand{\lamGB}{\lambda_\text{\tiny{GB}}}
\newcommand{\lamcross}{\lambda_\times}

\newcommand{\hatA}{\hat{A}}

\begin{document}


\title{Pole-skipping with finite-coupling corrections}
%
\author{Makoto Natsuume}
\email{makoto.natsuume@kek.jp}
\altaffiliation[Also at]{
Department of Particle and Nuclear Physics, 
SOKENDAI (The Graduate University for Advanced Studies), 1-1 Oho, 
Tsukuba, Ibaraki, 305-0801, Japan;
 Department of Physics Engineering, Mie University, 
 Tsu, 514-8507, Japan.}
\affiliation{KEK Theory Center, Institute of Particle and Nuclear Studies, 
High Energy Accelerator Research Organization,
Tsukuba, Ibaraki, 305-0801, Japan}
\author{Takashi Okamura}
\email{tokamura@kwansei.ac.jp}
\affiliation{Department of Physics, Kwansei Gakuin University,
Sanda, Hyogo, 669-1337, Japan}
\date{\today}
\begin{abstract}
Recently, it is shown that many Green's functions are not unique at special points in complex momentum space using AdS/CFT. This phenomenon is similar to the pole-skipping in holographic chaos, and the special points are typically located at $\omega_n = -(2\pi T)ni$ with appropriate values of complex wave number $q_n$. We study finite-coupling corrections to special points. As examples, we consider four-derivative corrections to gravitational perturbations and four-dimensional Maxwell perturbations. 
While $\omega_n$ is uncorrected, $q_n$ is corrected at finite coupling. Some special points disappear at particular values of higher-derivative couplings. Special point locations of the Maxwell scalar and vector modes are related to each other by the electromagnetic duality.
\end{abstract}


\maketitle

\section{Introduction}

The AdS/CFT duality or holography \cite{Maldacena:1997re,Witten:1998qj,Witten:1998zw,Gubser:1998bc} is a useful tool to study strongly-coupled systems (see, \eg, Refs.~\cite{CasalderreySolana:2011us,Natsuume:2014sfa,Ammon:2015wua,Zaanen:2015oix,Hartnoll:2016apf}). Recently, a number of papers appeared which study a new aspect of retarded Green's functions using AdS/CFT \cite{Grozdanov:2019uhi,Blake:2019otz,Natsuume:2019xcy}. 

According to these works, many Green's functions are not unique at ``special points" in complex momentum space $(\omega, q)$, where $\omega$ is frequency and $q$ is wave number. Such a phenomenon is collectively known as ``pole-skipping" \cite{Grozdanov:2017ajz,Blake:2018leo,Grozdanov:2018kkt,Natsuume:2019sfp} and was originally discussed in the context of holographic chaos \cite{Shenker:2013pqa,Roberts:2014isa,Shenker:2014cwa,Maldacena:2015waa}. 

Main results drawn from recent works are
\begin{itemize}
\item 
Various Green's functions exhibit this phenomenon. In addition to the gravitational sound mode (energy density correlators) which was originally discussed in holographic chaos, the bulk scalar field, the bulk Maxwell field (the current and charge correlators from the boundary point of view), the gravitational shear mode (momentum correlators) and the tensor mode show this behavior. 
\item 
There is a universality for $\omega$. In all examples, special points are located at Matsubara frequencies. Typically, they start from $\nw := \omega/(2\pi T) =-i $ and continue $\nw_n = -in $ for a positive integer $n$%
\footnote{One would include hydrodynamic poles as special points as well. }.
For the sound mode, special points start from $\nw_{-1} = +i$. It is argued that the $\nw_{-1} = +i$ special point is related to a chaotic behavior. On the other hand, the value of $q_n$ depends on the system. 
\end{itemize}

The appearance of Matsubara frequencies $\omega_n =-(2\pi T) ni$ is intriguing, but this is a strong coupling result. The appearance may be an artifact of the strong coupling limit. The purpose of this paper is to study finite-coupling corrections (higher-derivative corrections from the bulk point of view) to special points. 

For the gravitational sound mode, there is a special point in the upper-half $\omega$-plane, $\nw_{-1}=+i$. Higher-derivative corrections to the special point have been discussed in Ref.~\cite{Grozdanov:2018kkt}. 

It is argued that this special point is related to a chaotic behavior. It is conjectured that a holographic system has the maximum Lyapunov exponent $\lambda_L=2\pi T$ \cite{Maldacena:2015waa}. It is also argued that higher-derivative corrections do not change the Lyapunov exponent. Thus, higher-derivative corrections should not change the $\nw_{-1}=+i$ special point. Ref.~\cite{Grozdanov:2018kkt} confirms this in the context of pole-skipping. (The butterfly velocity or $q_{-1}$ gets corrections.) 

Higher-derivative corrections to special points have been studied to some extent for the sound mode but have not been explored for the other perturbations which exhibit the pole-skipping. We study higher-derivative corrections to these ``non-chaotic" special points. As examples, we consider four-derivative corrections to
\begin{itemize}
\item pure gravity,
\item Einstein-Maxwell theory (in the four-dimensional neutral background).
\end{itemize}
We study the corrections to the first few special points and its implications. The main purpose is to show the universality of $\nw_n=-in$. Higher-derivative corrections do not affect $\nw_n=-in$ but affect $\nq_n$. In addition,
\begin{itemize}
\item Special points may disappear at a particular coupling (\sect{disappearance}).
\item The four-dimensional Maxwell theory has the electromagnetic duality \cite{Herzog:2007ij,Myers:2010pk}. The duality has an interesting consequence to the pole-skipping: special point locations of the Maxwell scalar and vector modes are related to each other (\sect{EM_dual}). We also comment on the relation between this property and a previous observation in Ref.~\cite{WitczakKrempa:2012gn}.
\item Ref.~\cite{Blake:2019otz} introduced the notion of ``anomalous points," and we make a few remarks (\sect{anomalous}). 
\end{itemize}

\section{Pole-skipping}

In this section, we briefly review Refs.~\cite{Grozdanov:2019uhi,Blake:2019otz,Natsuume:2019xcy}. We use the incoming Eddington-Finkelstein (EF) coordinates. For the Schwarzschild-AdS$_{p+2}$ (SAdS$_{p+2}$) black hole, the metric is given by%
\footnote{We use upper-case Latin indices $M, N, \ldots$ for the $(p+2)$-dimensional bulk spacetime coordinates and use Greek indices $\mu, \nu, \ldots$ for the $(p+1)$-dimensional boundary coordinates. The boundary coordinates are written as  $x^\mu = (t, x^i) =(t, \vecx)=(t,x,y,z\cdots)$. 
}
\begin{subequations}
\begin{align}
ds_{p+2}^2 &= -F(r) dt^2 + \frac{dr^2}{F(r)} + r^2d\vecx_p^2~, \\
&= - F(r) dv^2 +2 dv dr + r^2d\vecx_p^2~, \\
F(r) &= r^2(1-r^{-p-1})~, 
\label{eq:sads}
\end{align}
\end{subequations}
using the tortoise coordinate $dr_*:=dr/F$ and $v=t+r_*$. For simplicity, we set the AdS radius $L=1$ and the horizon radius $r_0=1$. We consider the perturbations of the form
\begin{align}
\phi(r) \, e^{-i\omega v +iqx}~.
%
\end{align}
As usual, we impose the incoming-wave boundary condition at the horizon. 

\subsection{Power series expansion}\label{sec:pole-skipping}

As a typical example of special points, consider the field equation of the form
\begin{subequations}
\label{eq:eom}
\begin{align}
0 = \phi''+P(r)\phi'+Q(r)\phi~.
%
\end{align}
The horizon $r=1$ is a regular singularity, and $P$ and $Q$ are expanded as
\begin{align}
P &= \frac{P_{-1}}{r-1} + P_0 + P_1(r-1) + \cdots~, \\
Q &= \frac{Q_{-1}}{r-1} + Q_0 + Q_1(r-1) + \cdots~,
%
\end{align}
\end{subequations}
in the EF coordinates. One typically has $P_{-1}=1-i\nw$ and $Q_{-1}=Q_{-1}(\nw,\nq^2)$, where $\nw:=\omega/(2\pi T)$ and $\nq:=q/(2\pi T)$. The field equation takes this form, \eg,  for
\begin{itemize}
\item the bulk scalar field,
\item the bulk Maxwell field (scalar mode and vector mode),
\item the gravitational perturbations (tensor mode and shear mode).
\end{itemize}
We mainly focus on these perturbations, where field equations take the form \eqref{eq:eom}.

The solution can be written as a power series:
\begin{align}
\phi(r) = \sum_{n=0}\, \phi_n\, (r-1)^{n+\lambda}~.
%
\end{align}
Substituting this into the field equation, one obtains the indicial equation at the lowest order: 
\begin{align}
\lambda_1=0~, \quad \lambda_2 =1-P_{-1}= i\nw~.
%
\end{align}
The coefficient $\phi_n$ is obtained by a recursion relation. The $\lambda_1$ ($\lambda_2$)-mode represents the incoming (outgoing) mode, and we choose the incoming mode $\lambda=\lambda_1=0$ hereafter. In the incoming EF coordinates, the incoming wave is a Taylor series. The $\lambda_2$-mode is not a Taylor series for a generic $\nw$. 

The situation changes when $i\nw$ is a nonnegative integer. Then, the $\lambda_2$-mode is also a Taylor series naively. But $\lambda_1$ and $\lambda_2$ differ by an integer. In such a case, the smaller root fails to produce the independent solution since the recursion relation breaks down at some $\phi_n$. Instead, the second solution would contain a $\ln(r-1)$ term and is not regular at $r=1$. 

However, this log term disappears for special values of $\nq$. Therefore, one has two regular solutions at $i\nw_n=n$ with appropriate $\nq_n$. Such a point is called a ``special point" or a ``pole-skipping point."

In order to obtain $(\nw_n, \nq_n)$ systematically, write the rest of the field equation in a matrix form \cite{Blake:2019otz}:
\begin{align}
0 & = M\phi \\
& =\begin{pmatrix} 
    M_{11} & M_{12} & 0 & 0 & \cdots \\
    M_{21} & M_{22} & M_{23} & 0 & \cdots \\
    \cdots & \cdots & \cdots & \cdots & \cdots 
  \end{pmatrix}
  \begin{pmatrix} 
    \phi_0 \\ \phi_1 \\ \cdots
  \end{pmatrix}~.
\label{eq:recursion_matrix}
\end{align}
Here,
\begin{align}
M_{ij} =a_{ij} i\nw + b_{ij} \nq^2+ c_{ij}~. 
\label{eq:Mij}
\end{align}
In particular, $M_{n,n+1} = n(n-1+P_{-1})=n(n-i\nw)$. The matrix $\calM^{(n)}$ is obtained by keeping the first $n$ rows and $n$ columns of $M$. The special points at $i\nw_n=n$ are obtained from
\begin{align}
\det \calM^{(n)}(\nw_n,\nq_n) = 0~.
\label{eq:condition}
\end{align}
For example, consider the first row:
\begin{align}
0 = M_{11} \phi_0 + M_{12}\phi_1~.
%
\end{align}
Normally, this equation determines $\phi_1$ from $\phi_0$. However, when $M_{12}=M_{11}=0$, both $\phi_0$ and $\phi_1$ are free parameters. The former condition gives $M_{12}=P_{-1}=1-i\nw=0$. The latter condition is $M_{11}=Q_{-1}=0$. Since $Q_{-1}$ contains $\nq^2$, there are 2 solutions of $\nq$ and 2 special points. 

The horizon $r=1$ is a regular singularity, but the horizon becomes a regular point at $(\nw_1,\nq_1)$ because $P_{-1}=Q_{-1}=0$. Ref.~\cite{Natsuume:2019xcy} uses this criterion to find $(\nw_1,\nq_1)$. Also, $\lambda_2=1$ at $\nw_1$, so the extra regular solution is the ``outgoing" solution we did not select previously.

Similarly, when $M_{23}=\det \calM^{(2)}=0$, $\phi_0$ and $\phi_2$ become free parameters. The former condition gives $M_{23}=2(2-i\nw)=0$. The latter condition is a degree-4 polynomial in $\nq$ since $M_{ij}$ contains $\nq^2$. Thus, there are 4 solutions of $\nq$ and 4 special points. One gets 
\begin{align}
\det \calM^{(2)} &= Q_{-1}(Q_{-1}+P_0)-P_{-1}Q_0~.
%
\end{align}

As is clear from this analysis, the appearance of Matsubara frequencies $i\nw_n=n$ comes from $\lambda_2-\lambda_1=i\nw$, 
and this is the consequence of the field equation of the form \eqref{eq:eom}.
We pay attention to this point when we examine field equations with higher-derivative corrections.

\subsection{Nonuniqueness}

At a special point, the bulk solution is not unique in the sense that it is parametrized by $\phi_n/\phi_0$. This is also written by the ``slope dependence" $\dq/\dw$. Consider the $\phi_n$-equation
\begin{align}
\frac{1}{N^{(n)}(\nw)} \det \calM^{(n)}(\nw,\nq) \phi_0 + (n-i\nw) \phi_n = 0
%
\end{align}
and expand near the special point $\nw=\nw_n+\delta\nw$ and $\nq=\nq_n+\delta\nq$. The field equation becomes
\begin{align}
&\frac{1}{N^{(n)}(\nw_n)} \left\{ \del_\nq\det \calM^{(n)}(\nw_n,\nq_n) \delta\nq 
\right. \nonumber \\
&\left. 
+ \del_\nw \det \calM^{(n)}(\nw_n,\nq_n)\delta\nw \right\} \phi_0 - i\delta\nw \phi_n = 0~.
\label{eq:phi_n}
\end{align}
The solution depends on $\phi_n/\phi_0$ and this is written in terms of $\delta\nq/\delta\nw$ how one approaches the special point. 

As a result of nonuniqueness of the bulk solution, the boundary Green's function is not unique. Generically, one would write a Green's function as
\begin{align}
G^R (\omega,q) = \frac{b(\omega,q)}{a(\omega,q)}~.
\label{eq:2pt}
\end{align}
Near the special point, the Green's function takes the form
\begin{subequations}
\begin{align}
G^R &= \frac{ \delta\omega (\del_\omega b)_n + \delta q (\del_q b)_n +\cdots }{\delta\omega (\del_\omega a)_n + \delta q (\del_q a)_n +\cdots } 
\\
&= \frac{ (\del_\omega b)_n + \frac{\delta q}{\delta\omega} (\del_q b)_n +\cdots }{ (\del_\omega a)_n + \frac{\delta q}{\delta\omega} (\del_q a)_n +\cdots }~,
%
\end{align}
\end{subequations}
and the Green's function at the special point is not uniquely determined. Rather, it depends on the slope $\delta q/\delta\omega$
\footnote{We should point out that the slope dependence may not take the form $\delta q/\delta\omega$. In this paper, we consider the slope dependence in a broader sense. It is a little subtle how one writes the slope dependence or how one approaches special points. This is related to ``anomalous points" in Ref.~\cite{Blake:2019otz}. See \sect{anomalous} for more details.}.

\subsection{Tensor decomposition}

We consider Maxwell and gravitational perturbations of the form $e^{-i\omega v +iqx}$. We consider these perturbations in neutral backgrounds, so they do not couple to each other. 
The perturbations are decomposed under the transformation of boundary spatial coordinate $x^i$. 
The Maxwell perturbations $A_M$ are decomposed as
\begin{align}
\text{scalar mode: }& A_v~, A_x~, A_r~, 
\nonumber \\
\text{vector mode: }& A_y~.
\nonumber 
%
\end{align}
For example, the scalar mode transforms as scalar under the transformation. 
Similarly, the gravitational perturbations are decomposed as 
\begin{align}
\text{scalar mode (sound mode): }& h_{vv}~, h_{vr}~, h_{rr}~, 
\nonumber \\
& h_{vx}~, h_{rx}~, h_{xx}~, h_{yy}~, 
\nonumber \\
\text{vector mode (shear mode): }& h_{vy}~, h_{ry}, h_{xy}~, 
\nonumber \\
\text{tensor mode: }& h_{yz}~.
\nonumber
%
\end{align}
Normally, one fixes the gauge $A_r=h_{rM}=0$ and constructs gauge-invariant variables which are invariant under the residual gauge transformation. Instead, 
we carry out analysis in a fully gauge-invariant manner. See Appendix~A of Ref.~\cite{Natsuume:2019sfp} for details. 

For the Maxwell perturbations, the vector mode $A_y$ is gauge-invariant by itself. For the scalar mode, the gauge-invariant variables are
\begin{subequations}
\begin{align}
\mfA_v &= A_v + \frac{\omega}{q}A_x~, 
\label{eq:inv_At} \\
\mfA_r &= A_r - \frac{1}{iq}A_x'~.
\label{eq:inv_Ar}
\end{align}
\end{subequations}
For the gravitational perturbations, the tensor mode $h_{yz}$ is gauge-invariant by itself. For the shear mode, the gauge-invariant variables are
\begin{subequations}
\begin{align}
\mfh_{vy} &= h_{vy} +\frac{\omega}{q} h_{xy}~, \\
\mfh_{ry} &= h_{ry} -\frac{r^2}{iq}\left( \frac{ h_{xy} }{r^2} \right)'~.
%
\end{align}
\end{subequations}

\section{Pure gravity}

\subsection{Higher-derivative corrections}

Higher-derivative corrections have been widely discussed in AdS/CFT. See, \eg, Refs.~\cite{Buchel:2004di,Brigante:2007nu,Buchel:2008sh,Myers:2009ij,Cremonini:2009sy,Buchel:2009sk,Myers:2010pk,WitczakKrempa:2012gn}. AdS/CFT has two couplings, 't~Hooft coupling $\lambda$ and the number of colors $N_c$. The leading Einstein gravity results are the large-$N_c$ limit, \ie, $\lambda\to\infty, N_c\to\infty$. The $1/\lambda$-corrections correspond to higher-derivative corrections or $\alpha'$-corrections. The $1/N_c$-corrections correspond to string loop corrections or quantum gravity corrections. We focus on the former corrections since the latter is difficult to evaluate in general and little is known. 

From string theory point of view, the bulk action is an effective action expanded in the number of derivatives. Schematically,
\begin{align}
S = \int d^{p+2}x \sqrt{-g} \{ \calL_2+ \calL_4 +\cdots \}~,
%
\end{align}
where $\calL_i$ denotes $i$-derivative terms. $\calL_2$ is the leading order Lagrangian: for pure gravity, 
\begin{align}
\calL_2=R-2\Lambda~, \quad 2\Lambda =- \frac{p(p+1)}{L^2}~.
%
\end{align}
We focus on the first nontrivial corrections with four derivatives. In general, one should include all possible independent terms%
\footnote{Some constraints may restrict the form of the corrections. For example, for the \Nfour\ super-Yang-Mills theory, the first nontrivial corrections appear at $O(\alpha'^3)$ because of supersymmetry.}. For pure gravity,
\begin{align}
\calL_4 &= L^2 ( \alpha_1 R^2 + \alpha_2 R_{AB}R^{AB} + \alpha_3 R_{ABCD}R^{ABCD} )~,
\label{eq:pure_gravity}
\end{align}
where $\alpha_i \sim \alpha'/L^2 \ll 1$ (we set $L=1$ below). The values depend on the theory one considers, but we assume that such a theory exists. For example, heterotic string theory does contain such terms. Also, for pure gravity, these
are the only possible corrections at $O(\alpha')$. But in the presence of matter fields such as the Maxwell field and a scalar field, one should include all possible four-derivative terms (see next section for the Einstein-Maxwell theory.)

\subsection{Field redefinitions}\label{sec:redef}

For higher-derivative corrections, it is important to take field redefinitions into account. Many coefficients $\alpha_i$ are actually ambiguous due to field redefinitions \cite{Tseytlin:1986zz}. In the absence of an off-shell formalism, the effective action is derived from the string theory $S$-matrix (see, \eg, \cite{Gross:1986mw}), but the $S$-matrix does not change under field redefinitions. 

As a simple example, consider a pure scalar theory 
\begin{align}
\calL_2 = - \frac{1}{2}(\nabla\phi)^2~.
%
\end{align}
Assume that the scalar has a shift symmetry $\phi\to\phi+c$ so that it appears only as $\nabla\phi$ in the action. There are 3 four-derivative terms:
\begin{align}
\calL_4 &= \beta_1 (\nabla^2\phi)^2 + \beta_2 (\nabla\phi)^2\nabla^2\phi + \beta_3 (\nabla\phi)^4~.
%
\end{align}
But under the field redefinition,
\begin{align}
\phi &= \tilde{\phi} + c_1\nabla^2\tilde{\phi} + c_2(\nabla\tilde{\phi})^2 + O(\alpha'^2)~,
%
\end{align}
the leading term changes as
\begin{align}
- \frac{1}{2}(\nabla\phi)^2 &= - \frac{1}{2}(\nabla\tilde{\phi})^2 \nonumber \\
& + c_1 (\nabla^2\tilde{\phi})^2 + c_2 (\nabla\tilde{\phi})^2\nabla^2\tilde{\phi} + O(\alpha'^2)
%
\end{align}
so  $\beta_i$ change as 
\begin{align}
\tilde{\beta}_1 =\beta_1 + c_1~, \quad 
\tilde{\beta}_2 =\beta_2 + c_2~, \quad
\tilde{\beta}_3 =\beta_3~.
%
\end{align}
Note that the field redefinition changes $O(\alpha'^2)$ terms as well.

The effective action has ambiguities at higher order, but this does not affect on-shell physics. The field redefinition for example changes the metric but does not change dimensionless physical quantities. 
Then, what one should do is to eliminate ambiguous terms as many as possible. 

This of course simplifies analysis. But, more importantly, one should check whether any nontrivial term is left. If there were none, one would not obtain nontrivial results. Also, we consider linear perturbations in this paper. In such a case, some further terms may be dropped because they are higher order in perturbations. 

For the pure scalar theory, there are 3 four-derivative terms and 2 field redefinition parameters. This leaves one term $\beta_3$, but it involves 4 powers of perturbations (in the $\phi=0$ background), so no nontrivial term is left. Consequently, higher-derivative corrections are trivial for the pure scalar theory. Similarly, the pure Maxwell theory in a neutral background has no nontrivial correction (see \sect{EM}). 
For nontrivial corrections, we analyze pure gravity and the Einstein-Maxwell theory.

Finally, as mentioned above, the field redefinition changes higher order terms in $\alpha'$ as well, so the equivalence under the field redefinition holds only at $O(\alpha')$ perturbatively. For nonperturbative results in $\alpha'$, the equivalence holds only if one takes into account higher order terms in $\alpha'$. 

For pure gravity, under the field redefinition 
\begin{align}
g_{MN} &= \tilde{g}_{MN} + a_1\tilde{R}_{MN} + \tilde{g}_{MN} (a_3\tilde{R}+a_5) + O(\alpha'^2)~,
\nonumber 
%
\end{align}
$\alpha_i$ change as
\begin{subequations}
\begin{align}
\tilde{\alpha}_1 &= \alpha_1 + \frac{1}{2}a_1 + \frac{p}{2}a_3~, \\
\tilde{\alpha}_2 &= \alpha_2 - a_1~, \\
\tilde{\alpha}_3 &= \alpha_3~,
%
\end{align}
\end{subequations}
with the rescaled cosmological constant (\appen{redef_app}):
\begin{align}
\tilde{\Lambda} = \Lambda \left[ 1+\frac{p+2}{p}\{a_1+(p+2)a_3\} \Lambda \right]~.
%
\end{align}
So, one can set $\alpha_1=\alpha_2=0$. Another choice is the Gauss-Bonnet combination:
\begin{align}
\calL_4 = \alpha (R^2 - 4 R_{AB}R^{AB} + R_{ABCD}R^{ABCD})~.
%
\end{align}
It is convenient to set $\alpha=\lamGB/(p-1)(p-2)$. This combination is particularly useful because the field equation is at most second order in derivatives. We consider this Gauss-Bonnet correction below.

\subsection{Pole-skipping}\label{sec:GB_results}

For Gauss-Bonnet gravity, the field equation is given by
\begin{align}
0&= R_{MN} - \frac{1}{2}g_{MN}R + g_{MN}\Lambda 
\nonumber \\
&-\frac{\alpha}{2}g_{MN} (R^2 - 4 R_{AB}R^{AB} + R_{ABCD}R^{ABCD}) 
\nonumber \\
&+2\alpha ( RR_{MN}-2R_{MA}R_N^{~A}
\nonumber \\
&-2R_{MANB}R^{AB} + R_{MABC}R_N^{~ABC} )~.
%
\end{align}
The black hole background of Gauss-Bonnet gravity is obtained in Ref.~\cite{Cai:2001dz}. In the incoming EF coordinates, 
\begin{subequations}
\begin{align}
ds_{p+2}^2 &= -F(r) \NGB^2 dv^2 +2 \NGB dvdr+r^2d\vecx_p^2~, \\
F(r) &= \frac{r^2}{2\lamGB} \left\{ 1-\sqrt{ 1-4\lamGB \left(1-\frac{1}{r^{p+1}}\right) } \right\}~, \\
\NGB^2 &= \frac{1}{2}\left( 1+\sqrt{1-4\lamGB} \right) \sim 1-\lamGB~.
%
\end{align}
\end{subequations}
The constant $\NGB$ is chosen so that the boundary metric takes the form $ds^2=r^2(-dv^2+d\vecx_p^2)$. The Hawking temperature is given by 
\begin{align}
2\pi T = \NGB \frac{p+1}{2}r_0~,
%
\end{align}
where we restored the horizon radius $r_0$.
The other thermodynamic quantities are
\begin{subequations}
\begin{align}
s& = \frac{1}{4G}r_0^p~, \\
\varepsilon &= \NGB \frac{p}{16\pi G}r_0^{p+1}~.
%
\end{align}
\end{subequations}
These quantities can be obtained from the Euclidean computation \cite{Brigante:2007nu}. Alternatively, one can use the Wald formula and the first law $d\varepsilon =Tds$. The entropy obeys the area law for planar Gauss-Bonnet black holes even in the presence of higher-derivative corrections.

We consider the $p=3$ tensor perturbation of the form
\begin{align}
h_{yz} =: r^2 e^{-i\omega v+iqx} \phi (r)~.
%
\end{align}
(See \appen{GB_generic} for $p>3$ Gauss-Bonnet gravity.) When $\lamGB=0$, the tensor mode equation takes the form of a minimally-coupled massless scalar field. With the Gauss-Bonnet term, the tensor mode equation is rather lengthy, so we do not present it explicitly. But recall that special points $i\nw_n=n$ come from $\lambda_2-\lambda_1=i\nw$. 
In the EF coordinates, the roots $\lambda$ are obtained from the near-horizon limit of the $\phi''$ and $\phi'$ terms of the field equation. In this limit, 
\begin{align}
\phi'' + \frac{1-i\nw}{r-1}\phi' + \frac{(\cdots)}{r-1}\phi \sim 0~, \quad (r\to1)~,
\label{eq:eom_horizon}
\end{align}
where $\nw$ is normalized by $\alpha'$-corrected temperature. The field equation takes the same form as \eq{eom}. Thus, the indicial equation gives $(\lambda_1,\lambda_2)=(0, i\nw)$, and the higher-derivative correction does not affect $i\nw_n=n$. 

Following \sect{pole-skipping}, the first few special points are obtained from
\begin{subequations}
\begin{align}
0&= \det \calM^{(1)}(\nw_1) 
\nonumber \\
&= 
- (1+8 \lamGB) \qN^2 - \frac{1}{2} (3+8 \lamGB)~, 
\label{eq:M1_tensor3} \\
0&= \det\calM^{(2)}(\nw_2) 
\nonumber \\
&=
(1+8 \lamGB )^2 \qN^4 
+2 \left(3+40\lamGB -64 \lamGB ^2 \right) \qN^2 
\nonumber \\
&+ 6 (1-4 \lamGB ) (1+8 \lamGB)~,
%
\end{align}
\end{subequations}
where $\qN:=\NGB \nq$. One then obtains
\begin{subequations}
\begin{align}
i\nw_1=1~, \quad 
& \nq_1^2 = -\frac{3+8\lamGB}{2\NGB^2(1+8\lamGB)} \\
& \sim -\frac{3}{2}+\frac{13}{2}\lamGB~, \\
i\nw_2=2~, \quad 
& \nq_{2,1}^2 \sim -\sqrt{3}(\sqrt{3}+1) - \frac{10}{\sqrt{3}+1}\lamGB~, \\
& \nq_{2,2}^2 \sim -\sqrt{3}(\sqrt{3}-1) + \frac{10}{\sqrt{3}-1}\lamGB~.
%
\end{align}
\end{subequations}

We used the field redefinition to consider the Gauss-Bonnet combination. But as long as the results are expressed at $O(\alpha')$ perturbatively, special point locations do not change under the field redefinition. See \appen{redef_app} for the details. So, consider the generic curvature-squared theories \eqref{eq:pure_gravity}. For example, $i\nw_1=1$ special point is given by
\begin{align}
i\nw_1=1~, \quad 
& \nq_1^2 \sim -\frac{3}{2}+13\alpha_3+O(\alpha'^2)~.
%
\end{align}
We confirmed the result explicitly by analyzing special points for \eq{pure_gravity}.

For the shear mode, the field equations can be written as first-order differential equations of gauge-invariant variables. Schematically,
\begin{subequations}
\label{eq:eom_shear}
\begin{align}
%
0 &= \mfh_{vy}' + M_{vv} \mfh_{vy} + M_{vr}\mfh_{ry}~,
\label{eq:eom_vy} \\
0 &= \mfh_{ry}' + M_{rv} \mfh_{vy} + M_{rr}\mfh_{rr}~. 
\label{eq:eom_ry} 
%
\end{align}
\end{subequations}
In order to implement the method of \sect{pole-skipping}, use the master equation. Write \eq{eom_vy} as $\mfh_{ry}=\mfh_{ry}( \mfh_{vy}', \mfh_{vy})$ and substitute it into \eq{eom_ry}. One then obtains the master equation with the master variable $\mfh_{vy}$. The master equation takes the same form as \eq{eom}, so the higher-derivative correction does not affect $i\nw_n=n$. The first few special points of the shear mode are
\begin{subequations}
\begin{align}
i\nw_1=1~, \quad
& \nq_1^2  = \frac{3+8\lamGB}{2\NGB^2(1-4\lamGB)} \\
& \sim \frac{3}{2}+\frac{23}{2}\lamGB~, \\
i\nw_2=2~, \quad
& \nq_{2,1}^2 \sim -\sqrt{6} + (16-7\sqrt{6})\lamGB~, \\
& \nq_{2,2}^2 \sim +\sqrt{6} + (16+7\sqrt{6})\lamGB~.
%
\end{align}
\end{subequations}
In the $\lamGB\to0$ limit, these results coincide with known results.

For the sound mode, the pole-skipping analysis is a little intricate because the field equation does not always take the same form as \eq{eom}. In this sense, the sound mode is not our main concern, but for completeness and for its importance, we discuss it in \appen{sound}.

\subsection{Disappearance of special points}\label{sec:disappearance}

So far we discussed higher-derivative corrections in a perturbative framework and presented results to $O(\lamGB)$. Field redefinitions do not affect the results. But in this subsection, we go beyond the perturbative analysis and consider some particular values of $\lamGB$. 

Not all values of the coupling are allowed though. The consistency of the dual theory prevents the coupling from becoming very large.
As is clear from the metric, $\lamGB\leq1/4$, but there is a more stringent constraint from the causality of the dual theory \cite{Buchel:2009sk}:
\begin{align}
-\frac{(3p+5)(p-1)}{4(p+3)^2} \leq \lamGB \leq \frac{ (p^2+p+6)(p-1)(p-2) }{ 4(p^2-p+4)^2 }~.
\label{eq:bound_GB}
\end{align}
The upper bound reduces to $1/4$ in the $p\to\infty$ limit. For $p=3$, $-7/36 \leq \lamGB \leq 9/100$.

One should keep in mind that we truncate the action at $O(\alpha')$ here. When one considers particular values of $\lamGB$, one can no longer ignore the other higher-derivative terms at $O(\alpha'^2)$ and higher. Also, the equivalence under field redefinitions no longer holds. As mentioned in \sect{redef}, the field redefinition of the truncated action in general produces the other higher-derivative terms. Thus, statements as rigorous as the perturbative analysis are not possible here. One should regard the truncated action as a toy model. 

However, going beyond the perturbative analysis, one has a qualitatively new feature. Some special points ``disappear" at a particular $\lamGB$.

The first special point is determined by 
\begin{align}
0 = M_{11} \phi_0 + (1-i\nw)\phi_1~.
%
\end{align}
When $i\nw_1=1$ and $M_{11}=0$, both $\phi_0$ and $\phi_1$ become free parameters, and one has two regular solutions. The condition $M_{11}=0$ is satisfied by choosing an appropriate $\nq^2$. However, at finite coupling, we have one more parameter $\lamGB$. One can fine-tune $\lamGB$ so that the $\nq^2$-coefficient of $M_{11}$ vanishes. Then, $M_{11}\neq0$ and $\phi_0$ must vanish. As a result, there is a unique regular solution. In fact, $M_{11} = Q_{-1} \neq 0$, so the horizon remains a regular singularity: another solution should not  be regular, and one expects a $\ln(r-1)$ solution.

For the tensor mode, $M_{11} = \det \calM^{(1)}$ is given in \eq{M1_tensor3}, and the $\nq^2$-coefficient vanishes at 
\begin{align}
\lamGB=\lamcross=-\frac{1}{8}~.
%
\end{align}
The special point $\nw_1$ disappears there. This lies inside the bound \eqref{eq:bound_GB}. Since we use the truncated action, the precise value of $\lamcross$ is likely to change by the other higher-derivative corrections. 

The disappearance also affects the number of special points at $i\nw_n$. The pole-skipping condition is $\det \calM^{(n)}=0$. In general, this is a degree-$(2n)$ polynomial in $\nq$, which gives $(2n)$ special points. At the disappearance point $\lamcross$, $M_{11}$ is $\nq$-independent, so the degree of the pole-skipping condition decreases. For the tensor mode, the number of special points decreases as follows:
\begin{align}
&(i\nw_n, \text{number of } \nq_n) 
\nonumber \\
&= (1,0), (2,2), (3,2), (4,4), (5,4), (6,6), (7,6), \ldots
\nonumber
%
\end{align}

For the shear mode, the special point $\nw_1$ does not disappear inside the bound (\appen{expressions}). Actually, for theories considered in this paper, only the tensor mode with $p=3,4$ has a disappearance point inside bounds. The disappearance is a new interesting phenomenon, and it can occur in principle. But combined with such bounds, the disappearance does not seem to occur frequently. Special points may be protected from disappearance.

The disappearance is particularly interesting if it occurs in the sound mode of gravitational perturbations  because its special point is related to a chaotic behavior. The sound mode has the first special point at $\nw_{-1}=+i$ which reflects the maximum Lyapunov exponent $\lambda_L=2\pi T$. 

The higher-derivative correction to the $\nw_{-1}$ special point has been discussed in Ref.~\cite{Grozdanov:2018kkt} for the $p=3$ Gauss-Bonnet gravity (see \appen{sound} for $p\geq3$ Gauss-Bonnet gravity).
The special point is corrected as
\begin{align}
i\nw_{-1}=-1~, \quad
 \nq_{-1}^2 & = -\frac{2p}{(p+1)\NGB^2}~.
%
\end{align}
The result is valid nonperturbatively in $\lamGB$. 
Since $\NGB\neq0$, the disappearance of the special point does not occur in the sound mode of Gauss-Bonnet gravity. However, it would be interesting to examine whether the disappearance of the sound mode special point never occurs or not even if one uses the other higher-derivative corrections. Also, if it occurs, it would be interesting to study its implication to chaos. The out-of-time-ordered correlators (OTOC) are often used to study quantum many-body chaos, and it would be interesting to look at OTOCs at the disappearance point $\lamcross$.

\section{Einstein-Maxwell theory}\label{sec:EM}

\subsection{Higher-derivative corrections and field redefinitions}

In this section, we consider the four-dimensional Einstein-Maxwell theory:
\begin{align}
S = \int d^{4}x \sqrt{-g} \left[ R-2\Lambda-\frac{1}{4g_4^2}F^2 \right]~.
%
\end{align}
In the absence of sources, the four-dimensional bulk Maxwell theory is (Hodge) self-dual. From the boundary point of view, the duality is interpreted as ``particle-vortex" duality \cite{Herzog:2007ij}. As we see below, the self-duality has an interesting implication to the pole-skipping. 

Thus, we consider a neutral black hole background. Then, the Maxwell perturbations decouple from gravitational perturbations. The background is the SAdS$_4$ black hole:
\begin{subequations}
\begin{align}
ds_4^2 &= - F(r) dv^2 +2 dv dr + r^2d\vecx_2^2~, 
\label{eq:sads4} \\
F(r) &= r^2(1-r^{-3})~.
\end{align}
\end{subequations}
The Hawking temperature is given by $2\pi T=3/2$. 

Again we consider all possible four-derivative terms. In the Einstein-Maxwell theory, there are 6 new independent terms \cite{Natsuume:1994hd,Myers:2009ij,Myers:2010pk}:
\begin{align}
\calL_4 &= 
\alpha_1 R^2 + \alpha_2 R_{AB}R^{AB} + \alpha_3 R_{ABCD}R^{ABCD}
\nonumber \\
&+ \alpha_4 (\nabla_A F^{AC})(\nabla^B F_{BC}) + \alpha_5 F^4 + \alpha_6 (F^2)^2 
\nonumber \\
&+ \alpha_7 R^{ABCD} F_{AB}F_{CD} + \alpha_8 R^{AB}F_{AC}F_B^{~C} + \alpha_9 RF^2~,
\nonumber
%
\end{align}
where $F^4 := F^{AB}F_{BC}F^{CD}F_{DA}$.
First, we reduce the number of terms by field redefinitions. There are 6 new terms in the action and 3 new field redefinition parameters (\appen{redef_app}). This leaves 3 terms in the action: one can choose $\alpha_5, \alpha_6,$ and $\alpha_7$. Second, the Maxwell field has no background. $\alpha_5$ and $\alpha_6$ terms involve 4 powers of perturbations, so they do not contribute to linear perturbation problems. For the pure gravity part, the four-dimensional Gauss-Bonnet term is a total derivative, so one can ignore them. 

Therefore, we end up with the Maxwell theory with only one nontrivial correction (in the SAdS$_4$ background): 
\begin{align}
S &= \frac{1}{g_4^2}\int d^{4}x \sqrt{-g} \left[ -\frac{1}{4}F^2 + \alpha R^{ABCD} F_{AB}F_{CD} \right]~.
\label{eq:action_riemann}
\end{align}
Instead, one often uses 
\begin{align}
S &= \frac{1}{g_4^2}\int d^{4}x \sqrt{-g} \left[ -\frac{1}{4}F^2 + \gamma C^{ABCD} F_{AB}F_{CD} \right]~,
\label{eq:action_weyl}
\end{align}
where $C_{ABCD}$ is the Weyl tensor:
\begin{align}
C_{ABCD} &= R_{ABCD} - \frac{2}{p}(g_{A[C}R_{D]B}-g_{B[C}R_{D]A}) 
\nonumber \\
&+ \frac{2}{p(p+1)}R g_{A[C}g_{D]B}~.
%
\end{align}
This does not affect perturbative analysis because these two corrections are related by field redefinitions. Ref.~\cite{Myers:2010pk} introduces this higher-derivative correction to break the self-duality of the Maxwell theory. We consider how the correction affects special points of the Maxwell theory. 

It is convenient to write the action in a general form:
\begin{align}
S &= \int d^{4}x \sqrt{-g} \left[ -\frac{1}{8g_4^2}F_{AB} X^{ABCD} F_{CD} \right]~.
\label{eq:Maxwell_general}
\end{align}
Then, the Maxwell theory with the correction can be written as 
\begin{align}
X_{AB}^{~~~CD} = I_{AB}^{~~~CD} -8\gamma C_{AB}^{~~~CD}~,
%
\end{align}
where
\begin{align}
I_{AB}^{~~~CD} := \delta_A^{~C}\delta_B^{~D} - \delta_A^{~D}\delta_B^{~C}~.
%
\end{align}

\subsection{Pole-skipping}

The field equation is given by
\begin{align}
0=\nabla_A \left[ F^{AB}-4\gamma C^{ABCD}F_{CD} \right]~.
%
\end{align}
The field equation is at most second order in derivatives for Maxwell perturbations. We first consider the vector mode $A_y e^{-i\omega v+iqx}$. 
The special points $i\nw_n=n$ come from $\lambda_2-\lambda_1=i\nw$. 
In the EF coordinates, the roots $\lambda$ are obtained from  $A_y''$ and $A_y'$ terms of the field equation, so it is enough to focus on this part of the field equation. The vector mode equation can be written as 
\begin{subequations}
\label{eq:eom_Weyl} 
\begin{align}
0 &= [ FGA_y']'-2i\omega G A_y' + (\cdots)A_y~, 
\\
G &:= 1+4\gamma\left(1-\frac{F}{r^2}\right)~.
%
\end{align}
\end{subequations}
Near the horizon $r=1$, $F(r)\sim 4\pi T(r-1)$, and $G(r)\sim 1+4\gamma$ which is nonvanishing from \eq{bound_Weyl} below. So, the field equation is approximately given by
\begin{align}
A_y'' + \frac{1-i\nw}{r-1}A_y' + \frac{(\cdots)}{r-1}A_y \sim 0~, \quad (r\to1)~.
%
\end{align}
The field equation takes the same form as \eq{eom}. Thus, the correction $\gamma$ does not affect $i\nw_n=n$. 

Following \sect{pole-skipping}, the first few special points of the Maxwell vector mode are
\begin{subequations}
\begin{align}
i\nw_1=1~, \quad
 &\nq_1^2 \sim 8\gamma~, \\
i\nw_2=2~, \quad
 &\nq_{2,1}^2 \sim 32\gamma~, \\
 &\nq_{2,2}^2 \sim -\frac{8}{3}+O(\gamma^2)~.
%
\end{align}
\end{subequations}

The scalar mode can be analyzed in a manner similar to the shear mode in \sect{GB_results}. The first few special points are
\begin{subequations}
\begin{align}
i\nw_1=1~, \quad
 &\nq_1^2 \sim -8\gamma~, \\
i\nw_2=2~, \quad
 &\nq_{2,1}^2 \sim -32\gamma~, \\
 &\nq_{2,2}^2 \sim -\frac{8}{3}+O(\gamma^2)~.
%
\end{align}
\end{subequations}
In the $\gamma\to0$ limit, these results coincide with known results. 
Note that
\begin{itemize}
\item
When $\gamma=0$, the vector and scalar modes have special points at the same locations. 
\item 
To $O(\gamma)$, the scalar mode special points are obtained from the vector mode ones by $\gamma\to-\gamma$.
\end{itemize}

Just like pure gravity, one may consider particular values of $\gamma$. The dual theory respects causality \cite{Myers:2010pk} if
\begin{align}
|\gamma| \leq \frac{1}{12}~.
\label{eq:bound_Weyl}
\end{align}
For the Maxwell vector and scalar modes, the first special points $\nw_1$ do not disappear inside the bound  (\appen{expressions}).

\subsection{Electromagnetic duality}\label{sec:EM_dual}

The special point locations of the Maxwell vector and scalar modes are related to each other. This is understood from the duality of the four-dimensional bulk Maxwell theory. 

First, consider $\gamma=0$. The Maxwell theory
\begin{align}
d({}^\star F)=0~, \quad dF=0~,
%
\end{align}
is self-dual under the Hodge dual transformation, $\hat{F}={}^\star F$. 
Write the charge-charge correlator $G_{vv}$ and the current-current correlator $G_{yy}$ as
\begin{subequations}
\begin{align}
G_{vv} &= -\frac{q^2}{\sqrt{q^2-\omega^2}} K^L~, \\ 
G_{yy} &= \sqrt{q^2-\omega^2} K^T~.
%
\end{align}
\end{subequations}
As a result of the self-duality, $K^L$ and $K^T$ satisfy \cite{Herzog:2007ij}
\begin{align}
K^T(\omega,q) K^L(\omega,q) = 1~.
\label{eq:self-dual}
\end{align}
Ref.~\cite{Herzog:2007ij} uses this relation to derive the constant conductivity. When $q=0$, $K^T=K^L$ from spatial isotropy. Then, \eq{self-dual} implies $K^T(\omega,0)=K^L(\omega,0)=-1$. Thus, the conductivity is constant and is frequency-independent:
\begin{align}
\sigma(\omega) =- \frac{G_{yy}(\omega,0)}{i\omega} = -K^T(\omega,0) = 1~.
%
\end{align}

The self-duality has an interesting consequence to special point locations. Suppose that $K^T$ has a special point $(\omega_n, q_n)$ and is not unique there. In order to retain \eq{self-dual}, $K^L$ is not unique there as well. For example, Ref.~\cite{Natsuume:2019xcy} obtained the Green's functions at $(\omega_1, q_1)$ which satisfy \eq{self-dual}%
\footnote{The``self-energies" $\Pi$ used in Ref.~\cite{Natsuume:2019xcy} is related to $K$ as $\Pi=\sqrt{k^2}K$. While $\nq_1=0$, note that discussion here is different from the last paragraph one. In fact, $K^T (\nw_1,\nq_1)\neq K^L(\nw_1,\nq_1)$. We define the Green's function at the special point by the limit $\dw, \dq\to0$. What is really meaningful here is the $q\neq0$ expressions. }.  

When $\gamma\neq0$, the self-duality is lost. But one can still construct a dual theory and the correlators still satisfy some relations \cite{Myers:2010pk}. Add the following term in the action \eqref{eq:Maxwell_general}
\begin{align}
S' = \int d^4x \sqrt{-g} \frac{1}{2} \epsilon^{ABCD}\hat{A}_A \del_B F_{CD}~,
%
\end{align}
and perform the functional integrals over $F_{MN}$ and $\hatA_M$. Here, $\epsilon_{0123}=\sqrt{-g}$. The duality comes from the functional integrations in two different orders. Performing the integral over $\hatA_M$ gives the Bianchi identity $\epsilon^{ABCD} \del_B F_{CD}=0$, which implies $F_{MN}=\del_M A_{N} - \del_N A_{M}$. What remains is the standard Maxwell theory with functional integral over $A_M$.

Instead, if one integrates out $F_{MN}$ first, the resulting action is given by
\begin{align}
\hat{S} &= \int d^{4}x \sqrt{-g} \left[ -\frac{1}{8\hat{g}_4^2} \hat{F}_{AB} \hat{X}^{ABCD} \hat{F}_{CD} \right]~,
%
\end{align}
where $\hat{g}_4:=1/g_4$, and $\hat{F}_{MN}:=\del_M \hatA_{N} - \del_N \hatA_{M}$. Also, 
\begin{subequations}
\begin{align}
&\hat{X}_{AB}^{~~~CD} =-\frac{1}{4}\epsilon_{AB}^{~~~EF} (X^{-1})_{EF}^{~~~GH} \epsilon_{GH}^{~~~CD}~, \\
& \frac{1}{2}(X^{-1})_{AB}^{~~~CD} X_{CD}^{~~~EF} = I_{AB}^{~~~EF}~.
%
\end{align}
\end{subequations}
The correlators of the original theory and the dual theory satisfy 
\begin{align}
K^T(\omega,q) \hat{K}^L(\omega,q) = 1~.
\label{eq:self-dual2}
\end{align}

For the standard Maxwell theory, $\hat{X}_{AB}^{~~~CD} = I_{AB}^{~~~CD}$, so the theory is self-dual. 
When $\gamma$ is small, one can show that 
\begin{subequations}
\begin{align}
(X^{-1})_{AB}^{~~~CD} &= I_{AB}^{~~~CD} + 8\gamma C_{AB}^{~~~CD} + O(\gamma^2)~, \\
\hat{X}_{AB}^{~~~CD} &= (X^{-1})_{AB}^{~~~CD} + O(\gamma^2)~,
%
\end{align}
\end{subequations}
so the dual transformation maps $\gamma\to-\gamma$ to $O(\gamma)$. 
Then, from \eq{self-dual2}, $K^T$ and $\hat{K}^L$ have a special point at the same location $(\omega_n, q_n(\gamma))$. Because the dual transformation maps $\gamma\to-\gamma$, $K^L$ has a special point at $(\omega_n, q_n(-\gamma))$.

When $q=0$, \eq{self-dual2} implies that the conductivities of the dual theory pair are the inverse of each other:
\begin{align}
\sigma(\omega,\gamma) = \frac{1}{ \hat{\sigma}(\omega,\gamma)} \sim \frac{1}{ \sigma(\omega,-\gamma)}~.
%
\end{align}
So, the poles and zeros of $\sigma$ are interchanged in the dual theory. Ref.~\cite{WitczakKrempa:2012gn} studies these poles and zeros since they are equally important. In the limit $\gamma\to0$, the poles and the zeros approach each other in the complex $\omega$-plane. They ``annihilate" at Matsubara frequencies since $\sigma$ must be constant \cite{WitczakKrempa:2012gn}. In retrospect, what they observed is a precursor of the pole-skipping: they study the overlaps of poles and zeros. They do not see nonuniqueness however because they take $\nq=0$ first.

\subsection{Comments on ``anomalous points"}\label{sec:anomalous}

Ref.~\cite{Blake:2019otz} introduced the notion of ``anomalous points," and we make some remarks. At a special point, a Green's function is not unique, but at an anomalous point, the Green's function does not take the ``pole-skipping form," namely it is not written as $\dq/\dw$.

When $\gamma=0$, $\nq_1=0$ (and $\nq_{2,1}=0$). This is an example of anomalous points%
\footnote{We discuss only the $\nq_1=0$ example below, but a similar remark applies to the other anomalous points.}. 
But, first of all, the Green's function is {\it not unique} at $(\nw_1,\nq_1)$. Ref.~\cite{Natsuume:2019xcy} explicitly shows that the Green's function depends on $\dqs/\dw$. However, one would write the Green's function in terms of $\dq/\dw$ and assume a finite $\dq/\dw$. Then, 
\begin{align}
\dqs/\dw = 2\nq_1 \dq/\dw =0~,
%
\end{align}
and the slope dependence vanishes. Namely, whether a special point is anomalous or not is merely the matter of how one approaches the special point. 

Moreover, we saw that $\nq_1\neq0$ in the presence of the higher-derivative correction. While $\nq_1$ is an anomalous point in the large-$N_c$ limit, it is no longer true at finite coupling. At anomalous points, a Green's function is not written as $\dq/\dw$ but is not uniquely determined. In our opinion, it is not really necessary to distinguish anomalous points from the other special points. 

If one uses expressions of \sect{pole-skipping}, anomalous points satisfy both \eq{condition} and
\begin{align}
\del_\nq \det \calM^{(n)}(\nw_n,\nq_n) = 0~.
%
\end{align}
In such a case, the first term of \eq{phi_n} vanishes, so the solution does not depend on $\dq/\dw$. But one could equally expand the equation in terms of $\nq^2$ and may replace the first term by $\del_{\nq^2} \det \calM^{(n)}(\nw_n,\nq_n)\, \dqs$, which may not vanish. For example, for the Maxwell vector mode, $\det \calM^{(1)} =M_{11} \propto \nq^2$.

\section{More on the universality}

Many Green's functions are not unique at Matsubara frequencies, and we have shown that this is valid even at finite coupling, but our analysis is far from complete. 
If one focuses on the universality of $i\nw_n=n$, one can consult previous works on higher-derivative corrections.

One often uses the Schwarzschild coordinates, so note the relation between the EF coordinates and the Schwarzschild coordinates. In the EF coordinates, we consider the perturbation $e^{-i\omega v}\phi \sim e^{-i\omega t} (r-1)^{-i\nw/2}\phi $ and
\begin{align}
\text{incoming: } \phi &\sim 1~, \nonumber \\
\text{outgoing: } \phi &\sim (r-1)^{i\nw}~.
\nonumber
%
\end{align}
On the other hand, in the Schwarzschild coordinates, one sets $e^{-i\omega t}\phi$ and 
\begin{align}
\text{incoming: } \phi &\sim (r-1)^{-i\nw/2}~,  \nonumber \\
\text{outgoing: } \phi &\sim (r-1)^{i\nw/2}~.
\nonumber
%
\end{align}
Either way, a special point arises when $\lambda_2-\lambda_1=i\nw$ is a nonnegative integer, where $\nw$ is normalized by $\alpha'$-corrected temperature. Thus, in the Schwarzschild coordinates, special points $i\nw_n=n$ eventually come from the well-known results 
\begin{align}
\phi \sim (r-1)^{\pm i\nw/2}~.
\label{eq:near_horizon}
\end{align}

There is a large literature of higher-derivative corrections, and we list only a few. One can see $i\nw_n=n$ from the following works but cannot see how $\nq_n$ is corrected:
\begin{itemize}
\item
Ref.~\cite{Buchel:2008sh} considers  the \Nfour\ SYM which has $O(\alpha'^3)$ corrections and analyze the $p=3$ shear and sound modes. 
\item 
Ref.~\cite{Brigante:2007nu} considers the $p=3$ Gauss-Bonnet gravity and analyze the tensor, shear, and sound modes, and our result of the universality is implicitly known from this work. 
\item 
Ref.~\cite{Buchel:2009sk} considers Gauss-Bonnet gravity in arbitrary dimensions and analyze the tensor, shear, and sound modes. This reference provides the master equations for these modes. While the near-horizon behavior is not explicitly stated, one can show $\lambda_2-\lambda_1=i\nw$ from their formulae and can carry out the pole-skipping analysis. In \appen{GB_generic}, we list a first few special points. 
\item
Ref.~\cite{Myers:2010pk} considers the $p=2$ Einstein-Maxwell theory in a neutral black hole background and analyze the Maxwell vector perturbation, and our result of the universality is implicitly known from this work.
\item 
Refs.~\cite{Myers:2009ij,Cremonini:2009sy} consider the $p=3$ Einstein-Maxwell theory in a charged black hole background and analyze the tensor mode.
\end{itemize}

In the Schwarzschild coordinates, if the solution with exponent $-i\nw/2$ exists, the time-reversal symmetry of gravity guarantees the existence of the solution with exponent $+i\nw/2$. What is nontrivial is that the difference is an integer. 
It is useful to write the field equation in the form of Schr\"{o}dinger equation. Use the tortoise coordinate $r_*$ and define a new field $\phi=: G(r) \varphi$. By choosing $G(r)$ appropriately, the field equation becomes
\begin{align}
\del_*^2\varphi + V(r) \varphi = \omega^2\varphi~.
\label{eq:schrodinger}
\end{align}
Incidentally, one often uses this form to derive the bound on couplings such as Eqs.~\eqref{eq:bound_GB} and \eqref{eq:bound_Weyl}. The effective potential $V(r)$ typically behaves as $V\sim (r-1)$ near the horizon. Then, the near-horizon solution is 
\begin{align}
\varphi \sim e^{\pm i\omega r_*} \sim (r-1)^{\pm i\nw/2}~,
%
\end{align}
where $4\pi T r_* \sim \ln(r-1)$. Thus, the near-horizon behavior \eqref{eq:near_horizon} follows from the following assumptions: 
\begin{enumerate}
\item The background is static.
\item There exists a master field $\phi$ and its field equation takes the form \eqref{eq:schrodinger}.
\item $V\sim (r-1)$ as $r\to1$~.
\item $G(1)$ is constant. 
\end{enumerate}

We are unaware of any general theorem, but not all systems satisfy these assumptions. As an example, consider the Maxwell vector mode with $\gamma=-1/4$. 
The effective potential is given in Eqs.~(5.11)-(5.13) of Ref.~\cite{Myers:2010pk}. When $\gamma=-1/4$,  $V\sim\text{(constant)}$ which violates the above assumption. In the EF coordinates, the field equation does not take the same form as \eq{eom}. One can see this from \eq{eom_Weyl}. However, $\gamma=-1/4$ is outside the bound \eqref{eq:bound_Weyl}. Thus, a generic bulk system does not satisfy the universality. One may need to impose some additional inputs such as the causality of the dual theory.

\section*{Acknowledgments}


This research was supported in part by a Grant-in-Aid for Scientific Research (17K05427) from the Ministry of Education, Culture, Sports, Science and Technology, Japan. 


\appendix 

\section{Field redefinitions}\label{sec:redef_app}

%
%
%
%
%
%


Consider field redefinitions of the Einstein-Maxwell theory:
\begin{subequations}
\label{eq:redef}
\begin{align}
g_{MN} &= \tilde{g}_{MN} + \delta\tilde{g}_{MN} + O(\alpha'^2)~, \\
A_M &= \tilde{A}_M + \delta\tilde{A}_M + O(\alpha'^2)~,
%
\end{align}
where
\begin{align}
\delta g_{MN} &= a_1 R_{MN} + a_2 F_{MA}F_N^{~A} 
\nonumber \\
&+ g_{MN} (a_3 R +a_4 F^2 + a_5)~, \\
\delta A_M &= b_1 A_M + b_2 \nabla^A F_{AM}~.
%
\end{align}
\end{subequations}
Here, we include the rescaling of the metric ($a_5$) and the Maxwell field ($b_1$). For simplicity, we set $g_4=1$. 
Under the field redefinitions, $\alpha_i$ change as follows:
\begin{subequations}
\begin{align}
\tilde{\alpha}_1 &= \alpha_1 + \frac{1}{2}a_1 + \frac{p}{2}a_3~, \\
\tilde{\alpha}_2 &= \alpha_2 - a_1~, \\
\tilde{\alpha}_3 &= \alpha_3~, \\
\tilde{\alpha}_4 &= \alpha_4 + b_2~, \\
\tilde{\alpha}_5 &= \alpha_5 + \frac{1}{2}a_2~, \\
\tilde{\alpha}_6 &= \alpha_6 - \frac{1}{8}a_2 - \frac{p-2}{8}a_4~, \\
\tilde{\alpha}_7 &= \alpha_7~, \\
\tilde{\alpha}_8 &= \alpha_8 + \frac{1}{2}a_1 - a_2~, \\
\tilde{\alpha}_9 &= \alpha_9 - \frac{1}{8}a_1 + \frac{1}{2}a_2 - \frac{p-2}{8}a_3 + \frac{p}{2}a_4~.
%
\end{align}
\end{subequations}

Consider the terms with the Maxwell field. There are 6 terms in the action ($\alpha_4, \ldots, \alpha_9$) and 3  field redefinition parameters ($a_2,a_4,b_2$). This leaves 3 terms in the action. One is the $\alpha_7$ term which does not change under the redefinitions. The $\alpha_4$ term can always be eliminated by $b_2$. The choice of the other 2 terms has some freedom, but some combination is not possible to choose. For example, the $(\alpha_5, \alpha_8)$ pair cannot be eliminated simultaneously in general since $a_1$ is chosen to eliminate $\alpha_2$. Similarly, for $p=2$, it is not possible to eliminate the $(\alpha_5, \alpha_6)$ pair or $(\alpha_6, \alpha_8)$ pair. We choose to eliminate the $(\alpha_8, \alpha_9)$ pair. 

Alternatively, one can construct combinations of $\alpha_i$ that remain invariant under the field redefinitions. Namely, the field redefinition parameters can be eliminated by appropriate combinations of $\alpha_i$. In the Einstein-Maxwell theory, there are 9 $\alpha_i$ and 5 field redefinition parameters ($a_1,a_2,a_3,a_4,b_2$), so there are 4 invariant couplings. Two are $\alpha_3$ and $\alpha_7$, and the other two are
\begin{align}
& \alpha_5 + \frac{1}{4}\alpha_2 + \frac{1}{2}\alpha_8~, \\
& \alpha_6 + \frac{(p-2)^2}{4p^2}\alpha_1 + \frac{3p^2-12p+8}{16p^2}\alpha_2 
\nonumber \\
&+ \frac{3p-8}{8p}\alpha_8 +\frac{p-2}{2}\alpha_9~.
%
\end{align}
One would expect that physical results depend only on these combinations of couplings.

The field redefinitions also affect two-derivative terms as
\begin{align}
\calL_2 &= \left[1-\{a_1+(p+2)a_3\}\Lambda+\frac{p}{2}a_5 \right] R 
\nonumber \\
&+ \left[ 1+ \frac{p+2}{2}a_5 \right] (-2\Lambda)
\nonumber \\
&+ \left[ -\frac{1}{4} - \{a_2+(p+2)a_4 \}\Lambda - \frac{p-2}{8}a_5 - \frac{1}{2}b_1 \right] F^2~.
%
\end{align}
In order to keep the canonical normalization of $\calL_2$, choose rescaling parameters as
\begin{subequations}
\begin{align}
\frac{p}{2} a_5 &=\{a_1+(p+2)a_3\} \Lambda~, \\
b_1 &=-\frac{1}{2p} \{ (p-2)a_1+(p^2-4)a_3
\nonumber \\
&+4p(a_2+(p+2)a_4)\} \Lambda~.
%
\end{align}
\end{subequations}
Then, the rescaled cosmological constant becomes
\begin{align}
\tilde{\Lambda} = \Lambda \left[ 1+\frac{p+2}{p}\{a_1+(p+2)a_3\} \Lambda \right]~.
%
\end{align}

One expects that dimensionless physical quantities do not change at $O(\alpha')$ under field redefinitions. But it is not entirely obvious that special point locations $(\nw_n,\nq_n)$ do not depend on ``schemes." For our theories, this can be checked in a few ways. 

First, one can explicitly check this. For the $p=3$ tensor mode, we explicitly carry out analysis for generic curvature-squared theories \eqref{eq:pure_gravity}. For the $p=2$ Einstein-Maxwell theory, we explicitly carry out  analysis both for \eq{action_riemann} and for \eq{action_weyl}. 

Second, at a special point, a field $\phi$ is not unique. A field redefinition $\phi=\tilde{\phi}+\delta\tilde{\phi}$ subtracts $\delta\tilde{\phi}$ perturbatively in $O(\alpha')$ from the field. Under such a perturbative change, the nonuniqueness should remain. 

Third, for our theories,
\begin{subequations}
\label{eq:leading_EOM}
\begin{align}
R_{MN} &= \frac{2\Lambda}{p} g_{MN} + O(\alpha')~,\\
R &= \frac{p+2}{p} \Lambda + O(\alpha')~,\\
F_{MN} &= O(\alpha')~.
%
\end{align}
\end{subequations}
Thus, the field redefinition \eqref{eq:redef} is just an overall scaling at $O(\alpha')$%
\footnote{Note that the field redefinition is not equivalent to the Weyl scaling because one cannot use field equations inside the action. }. 
The scaling can be compensated by an isotropic scaling of $x^M$ and $L$, and the metric returns to the original one. Since the scaling is involved, $\omega$ and $T$ can scale in general, but dimensionless quantities such as $\omega/T$ do not change. 

%
%

\section{Sound mode analysis}\label{sec:sound}

The pole-skipping analysis is a little intricate for the sound mode%
\footnote{We use the master equation of Ref.~\cite{Buchel:2009sk} for sound mode analysis below.}, 
but first consider a generic $(\nw,\nq)$. In this case, the analysis is similar to the other cases, and one can locate special points in the lower-half $\omega$-plane. Again, the master equation takes the same form as \eq{eom}, so the higher-derivative correction does not affect $i\nw_n=n$ for $n>0$. The first few pole-skipping conditions are given in \appen{expressions} and \ref{sec:GB_generic}. In previous examples, the pole-skipping condition is a degree-$(2n)$ polynomial in $\nq$, but this is not the case for the sound mode. This is because \eq{Mij} does not hold to the sound mode. 

Thus, analysis of the lower-half $\omega$-plane is similar, but there must be a special point $\nw_{-1}$ in the sound mode. Partly because one uses the master equation in the method of Ref.~\cite{Blake:2019otz}, this special point must be examined separately. 

For a generic $(\nw,\nq)$, the above analysis is fine, but the denominators of pole-skipping conditions vanish when 
\begin{align}
\qN^2 = \frac{2p}{p+1}\nw^2~,
\label{eq:sound_condition}
\end{align}
and this case must be examined separately. This condition changes the near-horizon behavior of the master equation. 
The master equation does not take the same form as \eq{eom}. Instead,
\begin{align}
P_{-1}=-1-i\nw~, \quad Q_{-2}=1+i\nw~.
%
\end{align}
Note $Q_{-2}\neq0$. The indicial equation $\lambda(\lambda-1)+P_{-1}\lambda+Q_{-2}=0$ now gives
\begin{align}
\lambda_1=1~, \quad \lambda_2 =1+i\nw
%
\end{align}
instead of $(\lambda_1,\lambda_2)=(0,i\nw)$. Thus, in this case, one would expect two Taylor series solutions at $i\nw = -1, 0, +1, \ldots$. The $\nw=0$ case corresponds to the hydrodynamic pole. One can carry out the pole-skipping analysis with $\lambda_1$ or $\lambda_2$. Both roots produce matrix $M$ in the form of \eq{recursion_matrix}, where
\begin{align}
M_{n,n+1} = (n+\lambda)(n+\lambda-1) + (n+\lambda)P_{-1} + Q_{-2}~.
%
\end{align}

For the root $\lambda_1$, $M_{n,n+1}=n(n-i\nw)$, so the potential pole-skipping points are $i\nw_n=n$. But in this case, $\nw$ and $\nq$ are related, so it is not always possible to satisfy pole-skipping conditions. One can check that the pole-skipping condition is satisfied only for $n=1$. 

For the root $\lambda_2$, $M_{n,n+1}=n(n+i\nw)$, so the potential pole-skipping points are $i\nw_n=-n$. But Taylor series solutions are possible only for $n=1$. Also, the pole-skipping condition is satisfied for $n=1$. The $\nw_{-1}$ special point appears in this way. Imposing \eq{sound_condition} on these two cases, one obtains
\begin{subequations}
\begin{align}
 i\nw_{-1}=-1~, \quad
& \nq_{-1}^2  = -\frac{2p}{\NGB^2(p+1)}~, \\
i\nw_{1}=1~, \quad
& \nq_{1}^2  = -\frac{2p}{\NGB^2(p+1)}~, 
%
\end{align}
\end{subequations}

%
%
%
%


\section{Pole-skipping conditions}\label{sec:expressions}

\begin{itemize}
\item
The $p=3$ shear mode:
\begin{subequations}
\begin{align}
&\det \calM^{(1)}(\nw_1) 
\nonumber \\
&=  (1-4\lamGB) \qN^2 
- \frac{1}{2}(3+8\lamGB)~,
\label{eq:M1_shear3} \\
&\det\calM^{(2)}(\nw_2) 
\nonumber \\
&= (1-4\lamGB) \{ -(1-4\lamGB)\qN^4
\nonumber \\
&+32\lamGB \qN^2+6(1+8\lamGB) \}~.
%
\end{align}
\end{subequations}
The $O(\nq^2)$ term of \eq{M1_shear3} vanishes at $\lamcross=1/4$, and  $\det \calM^{(1)}=-5/2$. Actually, one can show that the $\nq^2$-dependence completely disappears from the field equations at $\lamcross=1/4$, but this is outside the bound \eqref{eq:bound_GB}, so we do not consider this case further. 

\item
The $p=3$ sound mode:
\begin{subequations}
\begin{align}
&\det \calM^{(1)}(\nw_1) 
\nonumber \\
&=\frac{ 4(1-8\lamGB)\qN^2(\qN^2-1)+3(3+8\lamGB) }{ 4(3+2\qN^2) }~,
\label{eq:M1_sound3} \\
&\det\calM^{(2)}(\nw_2) 
\nonumber \\
&= \frac{1}{4} (1-8\lamGB)^2 \qN^4-\frac{1}{2}(1-8\lamGB+64\lamGB^2) \qN^2
\nonumber \\
&+ \frac{3}{2}(1+4\lamGB-32\lamGB^2)~.
%
\end{align}
\end{subequations}
The $O(\nq^4)$ term of \eq{M1_sound3} vanishes at $\lamcross=1/8$, and  $\det \calM^{(1)}=3/(3+2\qN^2)$, but this is outside the bound \eqref{eq:bound_GB}.

\item
The Maxwell vector mode:
\begin{subequations}
\begin{align}
&\det \calM^{(1)}(\nw_1) = 
3\frac{ -(1-8\gamma)\nq^2 +8 \gamma }{ 4(1+4\gamma) }~,
\label{eq:M1_vector} \\
&\det\calM^{(2)}(\nw_2) =
\frac{1}{ 16 (1+4 \gamma )^2 }
\nonumber \\
&\times \{ 9 (1-8 \gamma )^2 \nq^4 
+ 24 \left( 1-28\gamma+16 \gamma ^2 \right) \nq^2 
\nonumber \\
&-768 \gamma  (1+\gamma) \}
%
\end{align}
\end{subequations}
The $O(\nq^2)$ term of \eq{M1_vector} vanishes at $\gamma_\times=1/8$, and $\det \calM^{(1)}=1/2$, but this is outside the bound \eqref{eq:bound_Weyl}. 

\item
The Maxwell scalar mode:
\begin{subequations}
\begin{align}
&\det \calM^{(1)}(\nw_1) = 
\frac{3 (1+4 \gamma)}{ 4(1-8\gamma) }\nq^2 + \frac{6 \gamma }{1+4 \gamma}~,  
\label{eq:M1_scalar} \\
&\det\calM^{(2)}(\nw_2) =
-\frac{9 (1+4\gamma)^2}{16 (1-8 \gamma)^2} \nq^4
\nonumber \\
&
+\frac{3 \left( -1 -20\gamma +80 \gamma ^2\right) }{2 (1-8 \gamma )^2} \nq^2
-\frac{48 \gamma (1+\gamma)}{(1+4 \gamma)^2}~.
%
\end{align}
\end{subequations}
The $O(\nq^2)$ term of \eq{M1_scalar} would vanish at $\gamma_\times=-1/4$, but this case is actually irrelevant. The near-horizon behavior of the field equation changes at $\gamma_\times$. The field equation does not take the same form as \eq{eom} because $Q_{-2}\neq0$. Since this $\gamma_\times$ is outside the bound \eqref{eq:bound_Weyl}, we do not consider this case further. 

\end{itemize}

\section{Gauss-Bonnet gravity in arbitrary dimensions}\label{sec:GB_generic}

Ref.~\cite{Buchel:2009sk} derived the master equations for the tensor, shear, and sound modes of Gauss-Bonnet gravity in arbitrary dimensions. The master equations for the tensor and shear modes take the same form as \eq{eom}, so the higher-derivative correction does not affect $i\nw_n=n$. 

Special points are obtained from $\det \calM^{(n)}(\nw_n,\nq_n) = 0$. For the tensor mode,
\begin{subequations}
\begin{align}
& \det \calM^{(1)}(\nw_1) = M_{11}(\nw_1) = (p+1)^2\frac{C_1}{C_2}~, 
\label{eq:M1_tensor} \\
C_1&= (p+1)\left\{ \frac{(2-p)(p-1)}{(p+1)^2} + 2\lamGB \frac{p-4}{p+1} 
+ 4 \lamGB^2 \right\} \qN^2 
\nonumber \\
&+2 (p-2)\left\{ \frac{(1-p)p}{(p+1)^2} + 2\lamGB \frac{p-2}{p+1} 
+ 4 \lamGB^2 \right\}
\label{eq:M1_tensor_num} \\
C_2 &= 8 (p-2) \{ - p+1+ 2 \lamGB  (p+1) \}~,
%
\end{align}
\end{subequations}
where $\qN:=\NGB \nq$. The expression of $\det \calM^{(2)}$ is rather lengthy, so we do not present it explicitly. To $O(\lamGB)$, the first few special points are
\begin{align*}
i\nw_1=1~, \quad
&\nq_1^2 \sim 
- \frac{2p}{p+1}
- \frac{ 2(p^3-3p^2-6p-8) }{ (p-2)(p-1)(p+1) }\lamGB~, 
\\
i\nw_2=2~, \quad
&\nq_{2,1}^2 \sim 
- \frac{4\sqrt{p}(\sqrt{p}+1)}{p+1} 
\\
&- \frac{4(p^2-p+4)}{ (p-2)(p+1)(\sqrt{p}+1) } \lamGB~, 
\\
&\nq_{2,2}^2 \sim 
- \frac{4\sqrt{p}(\sqrt{p}-1)}{p+1} 
\\
&+ \frac{4(p^2-p+4)}{ (p-2)(p+1)(\sqrt{p}-1) } \lamGB~.
%
\end{align*}
%
%
The first special points disappear when the $\nq^2$-coefficient of \eq{M1_tensor_num} vanishes. For $p>3$, the disappearance occurs at 
\begin{subequations}
\begin{align}
\lambda_{\times, 1} &= \frac{4-p-\sqrt{5p^2-20p+24}}{4(p+1)}~,\\
\lambda_{\times, 2} &= \frac{4-p+\sqrt{5p^2-20p+24}}{4(p+1)}~.
%
\end{align}
\end{subequations}
$\det \calM^{(1)}=1/2$ for both cases. For $p=3$, the only solution is $\lambda_{\times, 1}=-1/8$.
Comparing with the bound \eqref{eq:bound_GB}, $\lambda_{\times, 1}$ is inside the bound for $p=3,4$, but $\lambda_{\times, 2}$ is always outside the bound. 

For the shear mode, 
\begin{align}
&\det \calM^{(1)}(\nw_1) 
\nonumber \\
&=\frac{ (p+1) \{ -p+1 +2 \lamGB  (p+1) \} }{8 (p-1)} \qN^2
\nonumber \\
&+\frac{(1-p) p + 2 \lamGB \left(p^2-p-2\right)+4 \lamGB ^2 (p+1)^2 }{4 (1-p)+8 \lamGB  (p+1)}~.
\label{eq:M1_shear}
\end{align}
\pagebreak
Again, we do not present $\det \calM^{(2)}$ explicitly. To $O(\lamGB)$, the first few special points are
\begin{align*}
i\nw_1=1~, \quad
& \nq_1^2 \sim 
\frac{2p}{p+1} 
+ \frac{ 2(3p^2+5p+4) }{ (p-1)(p+1) }\lamGB~, 
\nonumber \\
i\nw_2=2~, \quad
& \nq_{2,1}^2 \sim 
- \frac{4 \sqrt{p(p-1)}}{p+1}
\nonumber \\
& + 8\lamGB \left\{ \frac{p+1}{p-1} - \frac{\sqrt{p(p-1)} (p^2+p+2) }{(p-1)^2 (p+1)} \right\}~, 
\nonumber \\
& \nq_{2,2}^2 \sim 
+ \frac{4 \sqrt{p(p-1)}}{p+1}
\nonumber \\
& + 8\lamGB \left\{ \frac{p+1}{p-1} + \frac{\sqrt{p(p-1)} (p^2+p+2) }{(p-1)^2 (p+1)} \right\}~.
\nonumber
%
\end{align*}
The $O(\nq^2)$ term of \eq{M1_shear} would vanish at
\begin{align}
\lamcross = \frac{p-1}{2(p+1)}~,
%
\end{align}
but this case is irrelevant. The near-horizon behavior of the field equation changes at $\lamcross$. The field equation does not take the same form as \eq{eom} because $Q_{-2}\neq0$. Since this $\lamcross$ is always outside the bound \eqref{eq:bound_GB}, we do not consider this case further. 

For the sound mode,
\begin{subequations}
\begin{align}
& \det \calM^{(1)}(\nw_1) = (p+1)^2 \frac{C_1}{C_2}~, 
\label{eq:M1_sound} \\
C_1&= -\left[ 12\left\{ (p+1)\lamGB-\frac{p}{4} \right\}^2 + \frac{p(p-4)}{4} \right] \qN^4
\nonumber \\
&+ 4p \left\{ \frac{p^2-3p+2}{(p+1)^2} - 2\lamGB \frac{2p-3}{p+1} 
+ 4\lamGB^2 \right\} \qN^2
\nonumber \\
&+4p^2 \left\{ \frac{p(1-p)}{(p+1)^2} + 2\lamGB \frac{p^2-p-2}{p+1} 
+ 4\lamGB^2 \right\} 
\label{eq:M1_sound_num} \\
C_2 &= 8p \{ (p+1)\qN^2+2p \} \{ 1-p+2(1+p)\lamGB \}~.
%
\end{align}
\end{subequations}
The $O(\nq^4)$ term of \eq{M1_sound_num} never vanishes for $p>4$. For $p=4$, it vanishes at $\lamcross=1/5$, and  $\det \calM^{(1)}=8/(8+5\qN^2)$, but this is outside the bound \eqref{eq:bound_GB}.



\begin{thebibliography}{}


\bibitem{Maldacena:1997re}
  J.~M.~Maldacena,
  ``The Large N limit of superconformal field theories and supergravity,''
  Int.\ J.\ Theor.\ Phys.\  {\bf 38} (1999) 1113
   [Adv.\ Theor.\ Math.\ Phys.\  {\bf 2} (1998) 231]
  [hep-th/9711200].

\bibitem{Witten:1998qj}
  E.~Witten,
  ``Anti-de Sitter space and holography,''
  Adv.\ Theor.\ Math.\ Phys.\  {\bf 2} (1998) 253
  [hep-th/9802150].

\bibitem{Witten:1998zw}
  E.~Witten,
  ``Anti-de Sitter space, thermal phase transition, and confinement in gauge theories,''
  Adv.\ Theor.\ Math.\ Phys.\  {\bf 2} (1998) 505
  [hep-th/9803131].
  
\bibitem{Gubser:1998bc}
  S.~S.~Gubser, I.~R.~Klebanov and A.~M.~Polyakov,
  ``Gauge theory correlators from noncritical string theory,''
  Phys.\ Lett.\ B {\bf 428} (1998) 105
  [hep-th/9802109].


\bibitem{CasalderreySolana:2011us}
  J.~Casalderrey-Solana, H.~Liu, D.~Mateos, K.~Rajagopal and U.~A.~Wiedemann,
  \textit{Gauge/String Duality, Hot QCD and Heavy Ion Collisions} (Cambridge Univ.\ Press, 2014)
  [arXiv:1101.0618 [hep-th]].

\bibitem{Natsuume:2014sfa}
  M.~Natsuume,
  \textit{AdS/CFT Duality User Guide},
  Lecture Notes in Physics Vol. 903 (Springer Japan, Tokyo, 2015) 
  [arXiv:1409.3575 [hep-th]].

\bibitem{Ammon:2015wua}
  M.~Ammon and J.~Erdmenger,
  \textit{Gauge/gravity duality : Foundations and applications}
  (Cambridge Univ.\ Press, 2015).

\bibitem{Zaanen:2015oix}
  J.~Zaanen, Y.~W.~Sun, Y.~Liu and K.~Schalm,
  \textit{Holographic Duality in Condensed Matter Physics}
  (Cambridge Univ.\ Press, 2015).

\bibitem{Hartnoll:2016apf}
  S.~A.~Hartnoll, A.~Lucas and S.~Sachdev,
  \textit{Holographic quantum matter}
  (The MIT Press, 2018) 
  [arXiv:1612.07324 [hep-th]].


\bibitem{Grozdanov:2019uhi}
  S.~Grozdanov, P.~K.~Kovtun, A.~O.~Starinets and P.~Tadi\'c,
  ``The complex life of hydrodynamic modes,''
  arXiv:1904.12862 [hep-th].

\bibitem{Blake:2019otz}
  M.~Blake, R.~A.~Davison and D.~Vegh,
  ``Horizon constraints on holographic Green's functions,''
  arXiv:1904.12883 [hep-th].

\bibitem{Natsuume:2019xcy}
  M.~Natsuume and T.~Okamura,
  ``Nonuniqueness of Green's functions at special points,''
  JHEP {\bf 1912} (2019) 139
  [arXiv:1905.12015 [hep-th]].


\bibitem{Grozdanov:2017ajz}
  S.~Grozdanov, K.~Schalm and V.~Scopelliti,
  ``Black hole scrambling from hydrodynamics,''
  Phys.\ Rev.\ Lett.\  {\bf 120} (2018) no.23,  231601
  [arXiv:1710.00921 [hep-th]].
  
\bibitem{Blake:2018leo}
  M.~Blake, R.~A.~Davison, S.~Grozdanov and H.~Liu,
  ``Many-body chaos and energy dynamics in holography,''
  JHEP {\bf 1810} (2018) 035
  [arXiv:1809.01169 [hep-th]].

\bibitem{Grozdanov:2018kkt}
  S.~Grozdanov,
  ``On the connection between hydrodynamics and quantum chaos in holographic theories with stringy corrections,''
  JHEP {\bf 1901} (2019) 048
  [arXiv:1811.09641 [hep-th]].

\bibitem{Natsuume:2019sfp}
  M.~Natsuume and T.~Okamura,
  ``Holographic chaos, pole-skipping, and regularity,''
  PTEP {\bf 2020} (2020) no.1, 013B07
  [arXiv:1905.12014 [hep-th]].


\bibitem{Shenker:2013pqa}
  S.~H.~Shenker and D.~Stanford,
  ``Black holes and the butterfly effect,''
  JHEP {\bf 1403} (2014) 067
  [arXiv:1306.0622 [hep-th]].

\bibitem{Roberts:2014isa}
  D.~A.~Roberts, D.~Stanford and L.~Susskind,
  ``Localized shocks,''
  JHEP {\bf 1503} (2015) 051
  [arXiv:1409.8180 [hep-th]].

\bibitem{Shenker:2014cwa}
  S.~H.~Shenker and D.~Stanford,
  ``Stringy effects in scrambling,''
  JHEP {\bf 1505} (2015) 132
  [arXiv:1412.6087 [hep-th]].

\bibitem{Maldacena:2015waa}
  J.~Maldacena, S.~H.~Shenker and D.~Stanford,
  ``A bound on chaos,''
  JHEP {\bf 1608} (2016) 106
  [arXiv:1503.01409 [hep-th]].
 

\bibitem{Herzog:2007ij}
  C.~P.~Herzog, P.~Kovtun, S.~Sachdev and D.~T.~Son,
  ``Quantum critical transport, duality, and M-theory,''
  Phys.\ Rev.\ D {\bf 75} (2007) 085020
  [hep-th/0701036].
 
\bibitem{Myers:2010pk}
  R.~C.~Myers, S.~Sachdev and A.~Singh,
  ``Holographic Quantum Critical Transport without Self-Duality,''
  Phys.\ Rev.\ D {\bf 83} (2011) 066017
  [arXiv:1010.0443 [hep-th]].

\bibitem{WitczakKrempa:2012gn}
  W.~Witczak-Krempa and S.~Sachdev,
  ``The quasi-normal modes of quantum criticality,''
  Phys.\ Rev.\ B {\bf 86} (2012) 235115
  [arXiv:1210.4166 [cond-mat.str-el]].
 

\bibitem{Buchel:2004di}
  A.~Buchel, J.~T.~Liu and A.~O.~Starinets,
  ``Coupling constant dependence of the shear viscosity in N=4 supersymmetric Yang-Mills theory,''
  Nucl.\ Phys.\ B {\bf 707} (2005) 56
  [hep-th/0406264].
   
\bibitem{Brigante:2007nu}
  M.~Brigante, H.~Liu, R.~C.~Myers, S.~Shenker and S.~Yaida,
  ``Viscosity Bound Violation in Higher Derivative Gravity,''
  Phys.\ Rev.\ D {\bf 77} (2008) 126006
  [arXiv:0712.0805 [hep-th]].

\bibitem{Buchel:2008sh}
  A.~Buchel,
  ``Resolving disagreement for eta/s in a CFT plasma at finite coupling,''
  Nucl.\ Phys.\ B {\bf 803} (2008) 166
  [arXiv:0805.2683 [hep-th]].

\bibitem{Myers:2009ij}
  R.~C.~Myers, M.~F.~Paulos and A.~Sinha,
  ``Holographic Hydrodynamics with a Chemical Potential,''
  JHEP {\bf 0906} (2009) 006
  [arXiv:0903.2834 [hep-th]].

\bibitem{Cremonini:2009sy}
  S.~Cremonini, K.~Hanaki, J.~T.~Liu and P.~Szepietowski,
  ``Higher derivative effects on eta/s at finite chemical potential,''
  Phys.\ Rev.\ D {\bf 80} (2009) 025002
  [arXiv:0903.3244 [hep-th]].

\bibitem{Buchel:2009sk}
  A.~Buchel, J.~Escobedo, R.~C.~Myers, M.~F.~Paulos, A.~Sinha and M.~Smolkin,
  ``Holographic GB gravity in arbitrary dimensions,''
  JHEP {\bf 1003} (2010) 111
  [arXiv:0911.4257 [hep-th]].


\bibitem{Tseytlin:1986zz}
  A.~A.~Tseytlin,
  ``Ambiguity in the Effective Action in String Theories,''
  Phys.\ Lett.\ B {\bf 176} (1986) 92.

\bibitem{Gross:1986mw}
  D.~J.~Gross and J.~H.~Sloan,
  ``The Quartic Effective Action for the Heterotic String,''
  Nucl.\ Phys.\ B {\bf 291} (1987) 41.
  

\bibitem{Cai:2001dz}
  R.~G.~Cai,
  ``Gauss-Bonnet black holes in AdS spaces,''
  Phys.\ Rev.\ D {\bf 65} (2002) 084014
  [hep-th/0109133].
  
\bibitem{Natsuume:1994hd}
  M.~Natsuume,
  ``Higher order correction to the Garfinkle-Horowitz-Strominger string black hole,''
  Phys.\ Rev.\ D {\bf 50} (1994) 3949
  [hep-th/9406079].
   
\end{thebibliography}
\end{document}